\documentclass[aps,prb,floatfix,reprint,showpacs,superscriptaddress,groupedaddress]{revtex4-1}
\usepackage{latexsym}
\usepackage{graphicx}
\usepackage{rotating}
\usepackage[normalem]{ulem}
\usepackage{hyperref}
\usepackage{amsmath,amssymb,amsfonts}
\usepackage{bm}
\usepackage{color}
\usepackage{soul}  

\usepackage{color}

\newcommand{\lyxmathsym}[1]{\ifmmode\begingroup\def\b@ld{bold}
	\text{\ifx\math@version\b@ld\bfseries\fi#1}\endgroup\else#1\fi}

\usepackage{amsfonts}
\usepackage{mathrsfs}

\hypersetup{colorlinks=true,urlcolor=blue,linkcolor=blue}

\begin{document}
\title{Controlling T$_{c}$ through band structure and correlation engineering in collapsed and uncollapsed phases of iron arsenides}
\author{Swagata Acharya}
\affiliation{ King's College London, Theory and Simulation of Condensed Matter,
              The Strand, WC2R 2LS London, UK}
\email{swagata.acharya@kcl.ac.uk}
\author{Dimitar Pashov}
\affiliation{ King's College London, Theory and Simulation of Condensed Matter,
              The Strand, WC2R 2LS London, UK}
\author{Francois Jamet}
\affiliation{ King's College London, Theory and Simulation of Condensed Matter,
	The Strand, WC2R 2LS London, UK}          
\author{Mark van Schilfgaarde}
\affiliation{ King's College London, Theory and Simulation of Condensed Matter,
              The Strand, WC2R 2LS London, UK}

\begin{abstract}
Recent observations of selective emergence (suppression) of superconductivity in the uncollapsed (collapsed) tetragonal
phase of LaFe$_{2}$As$_{2}$ has rekindled interest in understanding what features of the band structure control the
superconducting T$_{c}$.  We show that the proximity of the narrow Fe-d$_{xy}$ state to the Fermi energy emerges as the
primary factor. In the uncollapsed phase this state is at the Fermi energy, and is most strongly correlated and source 
of enhanced scattering in both single and two particle channels. The resulting intense and broad low energy spin fluctuations 
suppress magnetic ordering and simultaneously provide glue for Cooper pair formation.  In the
collapsed tetragonal phase, the d$_{xy}$ state is driven far below the Fermi energy, which suppresses the
low-energy scattering and blocks superconductivity.  A similar source of broad spin excitation appears in uncollapsed
and collapsed phases of CaFe$_{2}$As$_{2}$. This suggests controlling coherence provides a way to 
engineer T$_{c}$ in unconventional superconductors primarily mediated through spin fluctuations.

\end{abstract}

\maketitle

Through careful control of growth and annealing conditions, LaFe$_{2}$As$_{2}$ (LFA) can be grown in the tetragonal
phase with markedly longer $c$-axis than the value in the equilibrium ``collapsed'' tetragonal (CT) phase
($c{=}11.01$\,\AA).  The ``uncollapsed'' tetragonal phase (UT) has $c{=}11.73$\,\AA.  Moreover, the UT phase is shown to
superconduct at 12.1\,K, while the CT phase is not a superconductor~\cite{akira}.  A parallel
phenomenon was observed in undoped CaFe$_{2}$As$_{2}$ (CFA).  At room temperature, the equilibrium phase is UT, but it
was recently shown that a CT phase can be induced by quenching films grown at high temperature~\cite{chen2016}.  In this case, the
undoped CT phase superconducts with T$_{c}$=25\,K.  The UT phase does not exist at low temperature because CFA undergoes
a transition from tetragonal (I4/mmm) to orthorhombic (Fmmm) phase at 170\,K\cite{huang2008}, with a
concomitant transition to an ordered antiferromagnetic state~\cite{ni2008}.  It is also possible to induce a CT phase at low
temperature by applying pressure~\cite{nini,kreyssig}: superconductivity was reported with T$_{c}$ 12\,K at 0.3\,GPa. 
Taken together, these findings re-kindle the longstanding question as to whether universal band features can explain
unconventional superconductivity.

%

\begin{figure}[ht!]
\includegraphics[width=0.48\textwidth]{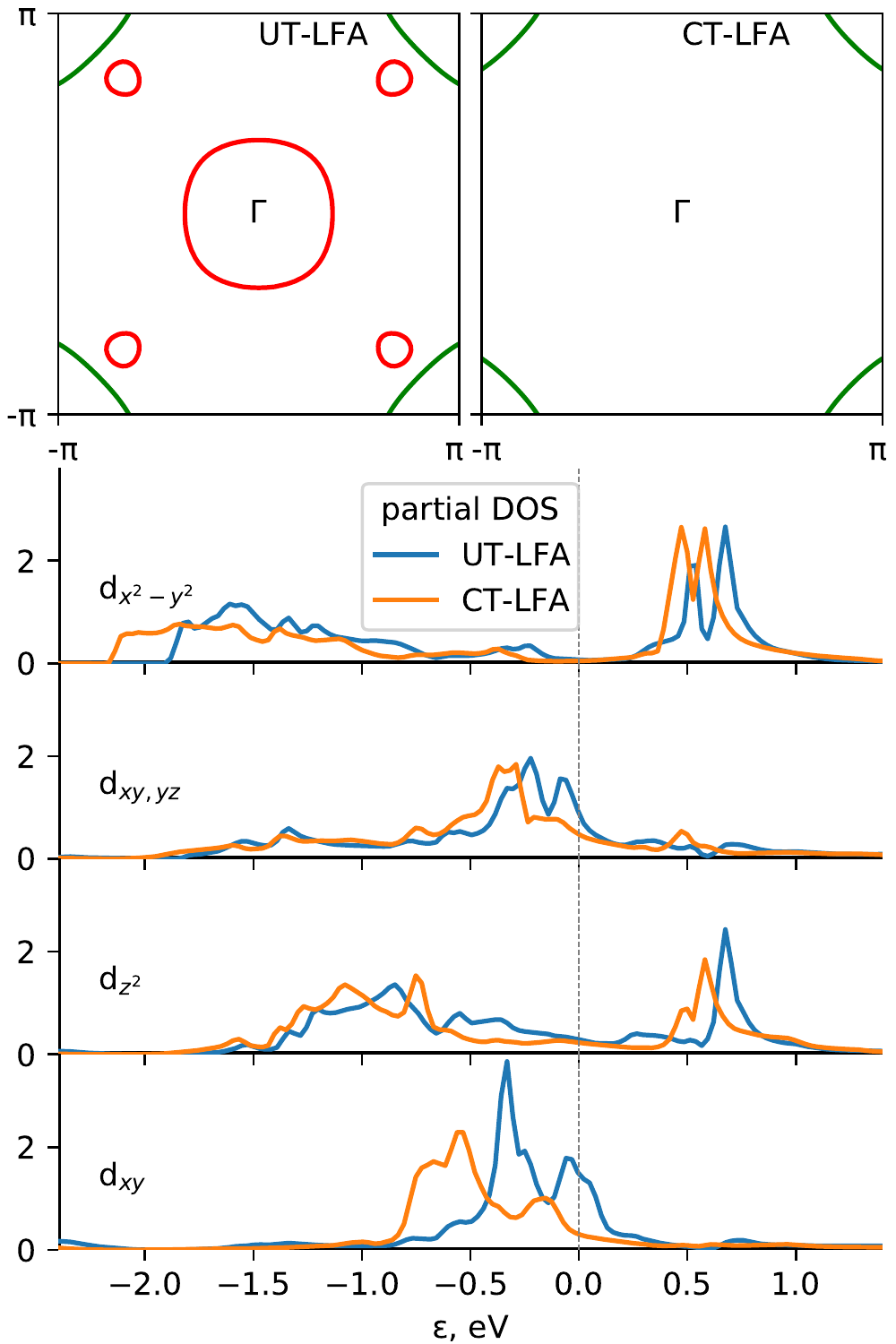}
\caption{Fermi surface in the k$_{z}$=0 plane for UT and CT-LFA phases.  In the UT phase, the circular pocket around
  $\Gamma$ has Fe-d$_{xy}$ character, while chickpea shaped pockets are of Fe-d$_{xz,yz}$ character. These pockets
  disappear in the CT phase where superconductivity is absent. Simultaneously, the effective band width $W$ of the Fe-3d
  manifold significantly increases in CT phase ($\sim$4~eV, in the UT phase $W{\sim}2.4$~eV, leading to larger
  electronic itineracy.  Also shown is the partial local density of states projected onto the Fe-3d orbitals. The bandwidth W of narrow d$_{xy}$ states gets further narrowed in UT-LFA phase to mark enhancement in effective correlation ($\frac{U}{W}$), where U is the Hubbard parameter.}
  \label{fig:fs}
\end{figure}

Here we use a recently developed \emph{ab initio} technique to show that there is indeed a universal feature, namely
incoherence originating from the Fe d$_{xy}$ state.  By `incoherence' we refer to the fuzzy spectral features and
momentum-broadened spin excitation caused by enhanced single- and two-particle scattering. Superconductivity depends
critically on the alignment of this state to the Fermi level.
We are able to make these findings thanks to recent developments that couple (quasi-particle) self consistent \emph{GW}
(QS\emph{GW}) with dynamical mean field theory (DMFT)~\cite{nickel,swag18,swag19,BaldiniPNAS}.  
Merging these two state-of-the-art
methods captures the effect of both strong local dynamic spin fluctuations (captured well in DMFT), and non-local
dynamic correlation~\cite{tomc, questaal-paper} effects captured by QS\emph{GW}~\cite{kotani}.  
We use QS\emph{GW} and not some other form of \emph{GW}, e.g. \emph{GW} based on DFT. It has been well established that QS\emph{GW} overcomes limitations of DFT-\emph{GW} when correlations become strong (see in particular Section 4 of Ref.~\cite{questaal-paper}).
On top of the DMFT
self-energy, charge and spin susceptibilities are obtained from vertex functions computed from the two-particle Green's
function generated in DMFT, via the solutions of non-local Bethe Salpeter equation.  Additionally, we compute the
particle-particle vertex functions and solve the linearized Eliashberg equation~\cite{hyowon_thesis,yin,swag19} to
compute the superconducting susceptibilities and eigenvalues of superconducting gap instabilities.  For CT and UT phases
we use a single value for \emph{U} and \emph{J} (3.5 eV and 0.62 eV respectively), which we obtained from bulk FeSe (and LiFeAs) within a constrained RPA
implementation following Ersoy et al.~\cite{christoph} DMFT is performed in the Fe-3d subspace, solved using a
rotationally invariant Coulomb interaction generated by these \emph{U} and \emph{J}.  The full implementation of the
four-tier process (QS\emph{GW}, DMFT, BSE, and BSE-SC) is discussed in Pashov et. al.~\cite{questaal-paper}, and codes are
available on the open source electron structure suite Questaal~\cite{questaal_web}.  Expressions we use for the response functions
are presented in Ref.~\cite{swag19}. Our all-electronic $GW$ implementation was adapted from the original ecalj package~\cite{ecalj};
the method and basis set are described in detail in Ref. ~\cite{questaal-paper}.
For the one-body part a $k$-mesh of $12\times12\times12$ was used; to compute the (much more weakly $k$-dependent) self-energy, we used a mesh of $6\times6\times6$ divisions, employing the tetrahedron method for the susceptibility.

We perform calculations in the tetragonal phases of  LFA and CFA; in the CT phase (CT-LFA and CT-CFA) and the corresponding
UT phase (UT-LFA and UT-CFA). Structural parameters for each phase are given in the SM, Table 1. The DMFT
self-energy, spin and charge susceptibilities, and finally the superconducting instability are computed as a function
of temperature.  CT-QMC samples more electronic diagrams at reduced
temperature and provides insights into the emerging coherence/incoherence in single- and two-particle instabilities;
however, it cannot provide knowledge about entrant structural (or structural+magnetic) transitions.  On the other hand,
it can tell us what would happen if the structural+magnetic transition could be suppressed (T$_{N}$=170\,K in CFA), and
we can estimate T$_{c}$ in the hypothetical UT phase of undoped CFA below T$_{N}$.

In brief, we find that the CT-LFA has no superconducting instability, while  UT, CT-CFA and UT-LFA are all
predicted to be superconducting. All of these findings are consistent with
experiment. In the experimentally known cases where the systems do superconduct (UT-LFA and CT-CFA), it
  appears our estimated T$_{c}$'s are a factor of two to three times larger than the experimental T$_{c}$. A similar
  discrepancy is observed in estimation of T$_{c}$ in doped single-band Hubbard model~\cite{hyowon_thesis}, where it
  sources from the local approximations of DMFT and needs a better momentum dependent vertex to circumvent
  this~\cite{kitatani}. Apart from a constant scaling, all of these findings are consistent with
experiment.  Moreover, we find that the hypothetical UT-CFA phase can have the  
highest T$_{c}$ of all.  We conclude that UT-CFA would be superconducting if it did not make a transition
to an antiferromagnetically ordered state.  The superior quality of the QS\emph{GW} bath combined with nonperturbative
DMFT has been shown to possess a high degree of predictive power in one- and two-particle spectral
functions~\cite{nickel,swag18,questaal-paper,swag19} and as in other cases we are able to replicate the experimental
observations of spectral functions, including a reasonable estimate for $T_{c}$.  The remainder of the paper uses this machinery
to explain what the origins of superconductivity are.

\begin{figure}[ht!]
	\includegraphics[width=0.48\textwidth]{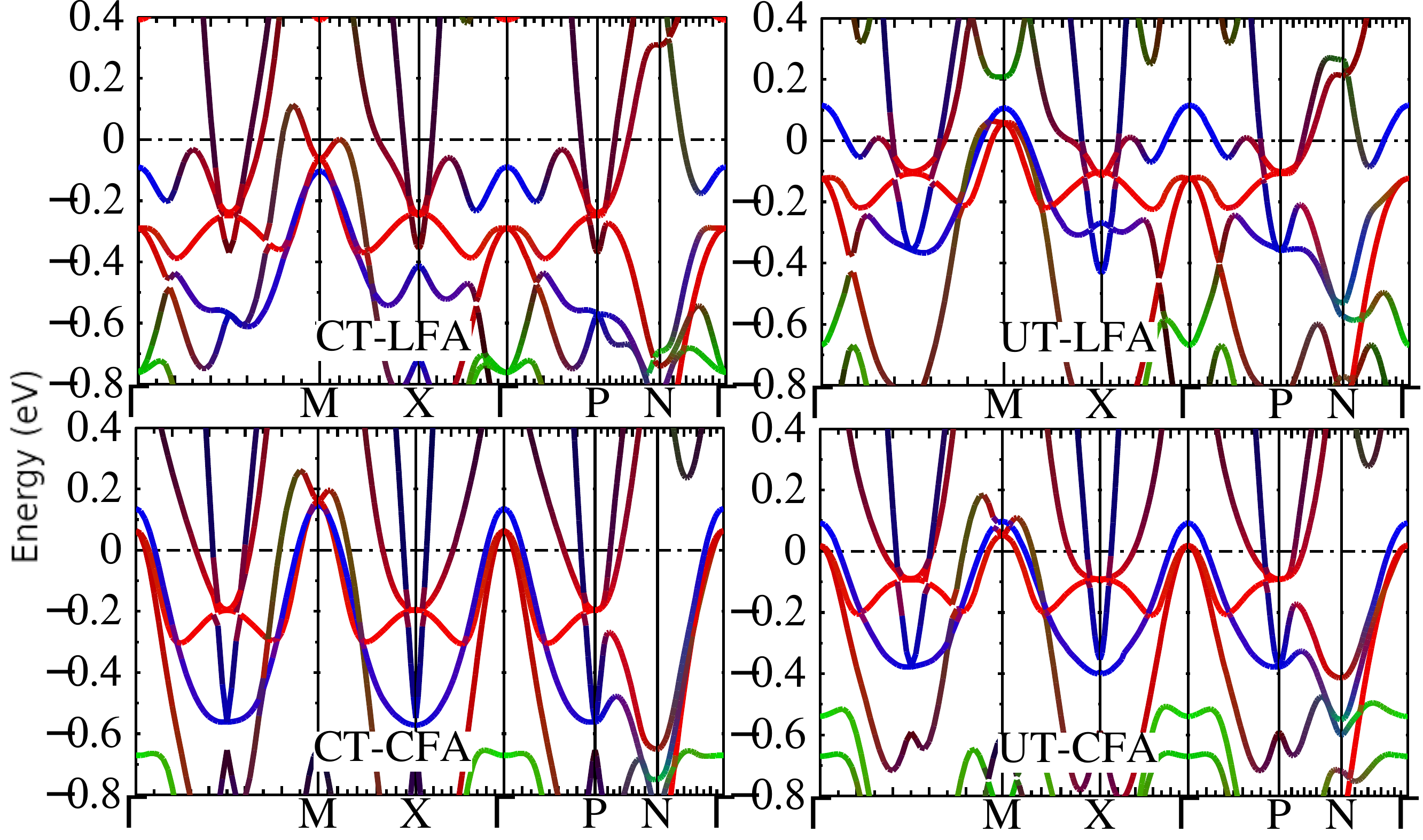}
	\caption{Color-weighted electronic QS\emph{GW} band structures in LFA (top) and CFA (bottom). CT (UT) phases are
		displayed on the left (right).  All t$_{2g}$ Fe states (d$_{xy}$ in blue, d$_{xz,yz}$ in red) are in close proximity
		to the Fermi energy, while the Fe-d$_{eg}$ states (depicted in green) are somewhat below.  In the CT phase of LFA,
		the d$_{xy}$ state is pushed below $E_{F}$, eliminating the hole
		pocket at $\Gamma$ and suppressing T$_{c}$.}
	\label{fig:band}
\end{figure}

The three systems predicted to have non-negligible T$_{c}$ (CT-CFA, UT-LFA, UT-CFA) have two things in common.  First,
the Fe-d$_{xy}$ state contributes to the hole pocket around the $\Gamma$ point (Fermi surface is shown in
Fig.~\ref{fig:fs}; see also the blue band in Fig.~\ref{fig:band}). Second, the imaginary part of the spin susceptibility
Im\,$\chi(q,\omega)$ has intense peaks centered at $\mathbf{q}{=}(\frac{1}{2},\frac{1}{2}$,0)$2\pi/a$, in the energy
window (2,25)\,meV.  The latter is a consequence of the former: low-energy spin-flip transitions involving d$_{xy}$ are
accessible, which give rise to strong peaks in Im\,$\chi(q,\omega)$ around the antiferromagnetic nesting vector
$\mathbf{q}^\mathrm{AFM}{=}(\pi/a,\pi/a,0)$.  Im\,$\chi(q,\omega)$ is diffused in $\mathbf{q}$ around
$\mathbf{q}^\mathrm{AFM}$.  This broadening in momentum space suppresses antiferromagnetism to allow superconductivity
to form.
CT-LFA is the only one of the four systems that has negligible instability to superconductivity.  In CT-LFA the
Fe-d$_{xy}$ state is pushed down (Fig.~\ref{fig:band}).  As a consequence the peak in
Im\,$\chi(\mathbf{q}^\mathrm{AFM},\omega)$ occurs at a much higher energy --- too high to provide the low-energy glue for Cooper
pairs.  Also appearing is a pronounced dispersive paramagnon branch around $q{=}0$.  This branch is present in all four
systems, but it is strongest in CT-LFA.  Nevertheless the \emph{ab initio} calculations predict no superconductivity.
This establishes that the paramagnon branch contributes little to the glue for superconductivity in these 122-As based compounds. 

\begin{figure}[ht!]
\includegraphics[width=0.5\textwidth]{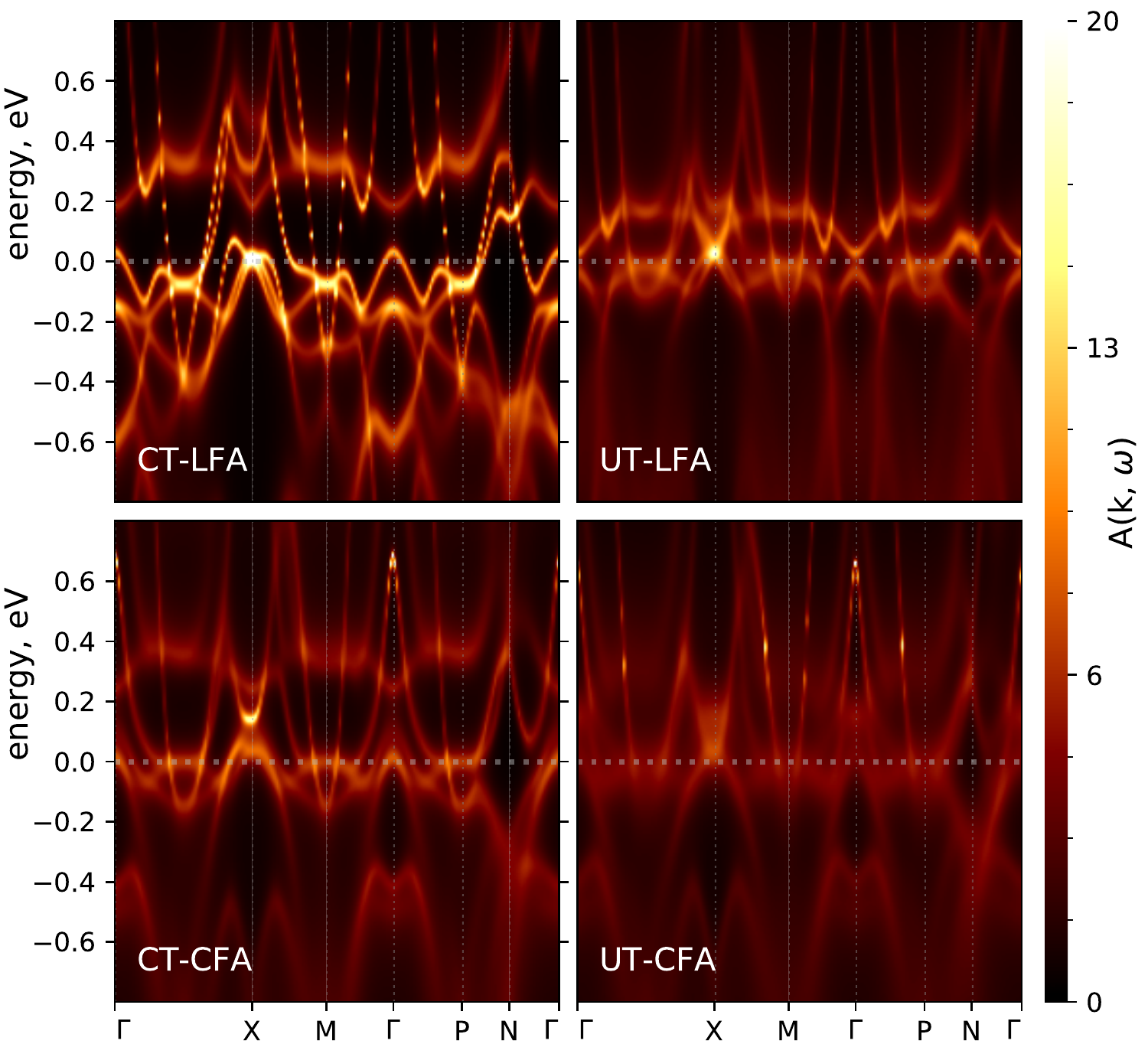}
\caption{Single-particle correlated QS\emph{GW}+DMFT electronic spectral functions $A(q,\omega)$ for CT and UT phases of
  LF and CFA along high-symmetry lines. The UT-CFA phase is most incoherent, while the CT-LFA phase is most coherent. The
  presence (absence) of Fe-d$_{xy}$ state at Fermi energy appears to be the primary criterion for incoherent (coherent)
  spectral features.}
\label{fig:specf}
\end{figure}

Reducing the $c$-axis in LFA phase pushes d$_{xy}$ below the Fermi energy $E_{F}$ (top left panel, Fig ~\ref{fig:band}); the
remaining hole pocket at $\Gamma$ is without d$_{xy}$ character (see Fig ~\ref{fig:fs}).  Quasi-particles in CT-LFA are
much more coherent (see Fig ~\ref{fig:specf}) with small scattering rate $\Gamma$ (extracted from the imaginary part of
the self-energy at $\omega{\rightarrow}0$) and large quasi-particle weights $Z$ relative to the other cases (see SM
Table~2 for the orbitally resolved numbers).  This further confirms that the CT-LFA phase is itinerant with
small correlation, using $U/W$ as a measure.
When the d$_{xy}$ state crosses $E_{F}$, single-particle spectral functions $A(q,\omega)$ become markedly incoherent.
This originates from enhanced single-particle scattering induced by local moment fluctuations within DMFT and suppressed
orbitally resolved $Z$ (SM Table 2). In the superconducting cases the d$_{xy}$ orbital character is the primary source
of incoherence with high scattering rate ($\Gamma{>}60$~meV) and quasi-particle weight as low as $\sim 0.4$.

\begin{figure}[ht!]
\includegraphics[width=0.50\textwidth]{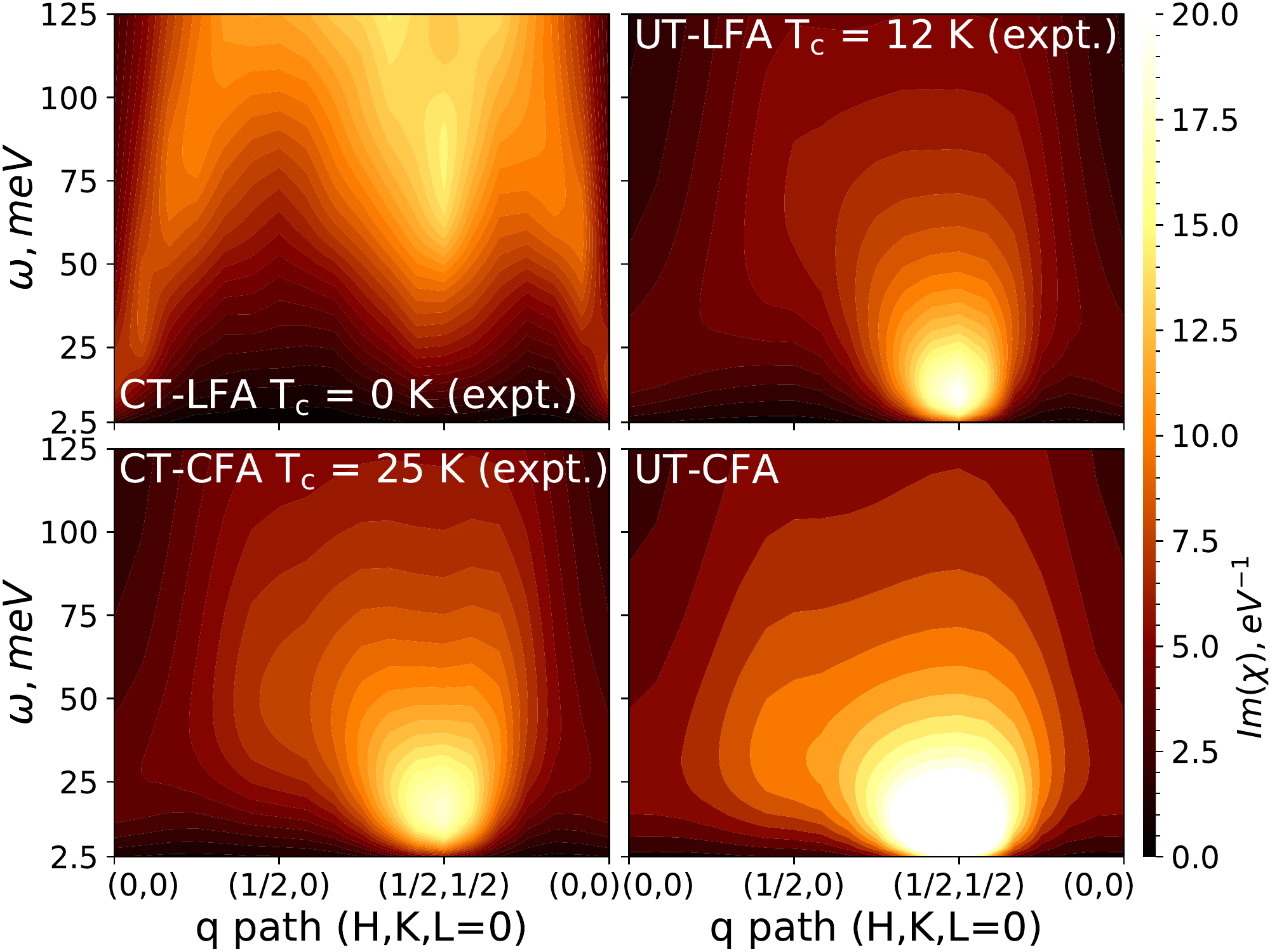}
\caption{The energy and momentum resolved spin susceptibility Im$\chi(q,\omega)$ (in the top panel from left to right)
  shown for the CT and UT phases respectively . The q-path (H,K,L=0) is chosen along
  (0,0)-($\frac{1}{2},0$)-($\frac{1}{2},\frac{1}{2}$)-(0,0) in the Brillouin zone corresponding to the two-Fe atom unit
  cell. The intensity in the CT phase is artificially multiplied by five to bring excitations for CT and UT phases to
  the same scale.}
\label{fig:chi}
\end{figure}

The peak in Im\,$\chi(\mathbf{q}^\mathrm{AFM},\omega)$
can be observed in almost all iron based superconductors~\cite{yin,yin2011}.
However, what varies significantly over
various systems is the dispersion of the branches. 
The less itinerant the system is, the smaller dispersion in Im$\chi(q,\omega)$ (and typical spin exchange scale $J \sim
t^2/U$), and it is more strongly correlated.

In the UT-LFA phase, Im$\chi(q,\omega)$ has a dispersive magnon branch extending to $\sim$70 meV. As can be observed in
Fig.~\ref{fig:chi}, both the branch and the low-energy peak at ($\frac{1}{2},\frac{1}{2}$,0) are significantly broad.  The
dispersion is significantly smaller than in undoped BaFe$_{2}$As$_{2}$ (BFA)~\cite{bfa-ins}, dispersion survives up to
200 meV at ($\frac{1}{2},0,0)$.  This suppression of branches and concomitant broadening suggests that UT-LFA is more
correlated than BFA.
In contrast with UT-LFA, CT-LFA has Stoner like continuum of spin excitations (in the figure the intensity is scaled
by a factor of five to make it similar to the UT phase) without any well defined low energy peak.  Similar
spin excitations can be observed in the phosphorus compounds (BaFe$_{2}$P$_{2}$, LiFeP) where the system either does
not superconduct or T$_{c}$ is fairly low (when it does)~\cite{yin}.  These are among the most itinerant systems of all iron based
superconductors and both the quasi-particle and spin excitations are band like.  In both the phases we find weak to no
$q_z$-dispersion of the susceptibilities, making the spin fluctuations effectively two dimensional.

\begin{figure}[ht!]
	\includegraphics[width=0.49\textwidth]{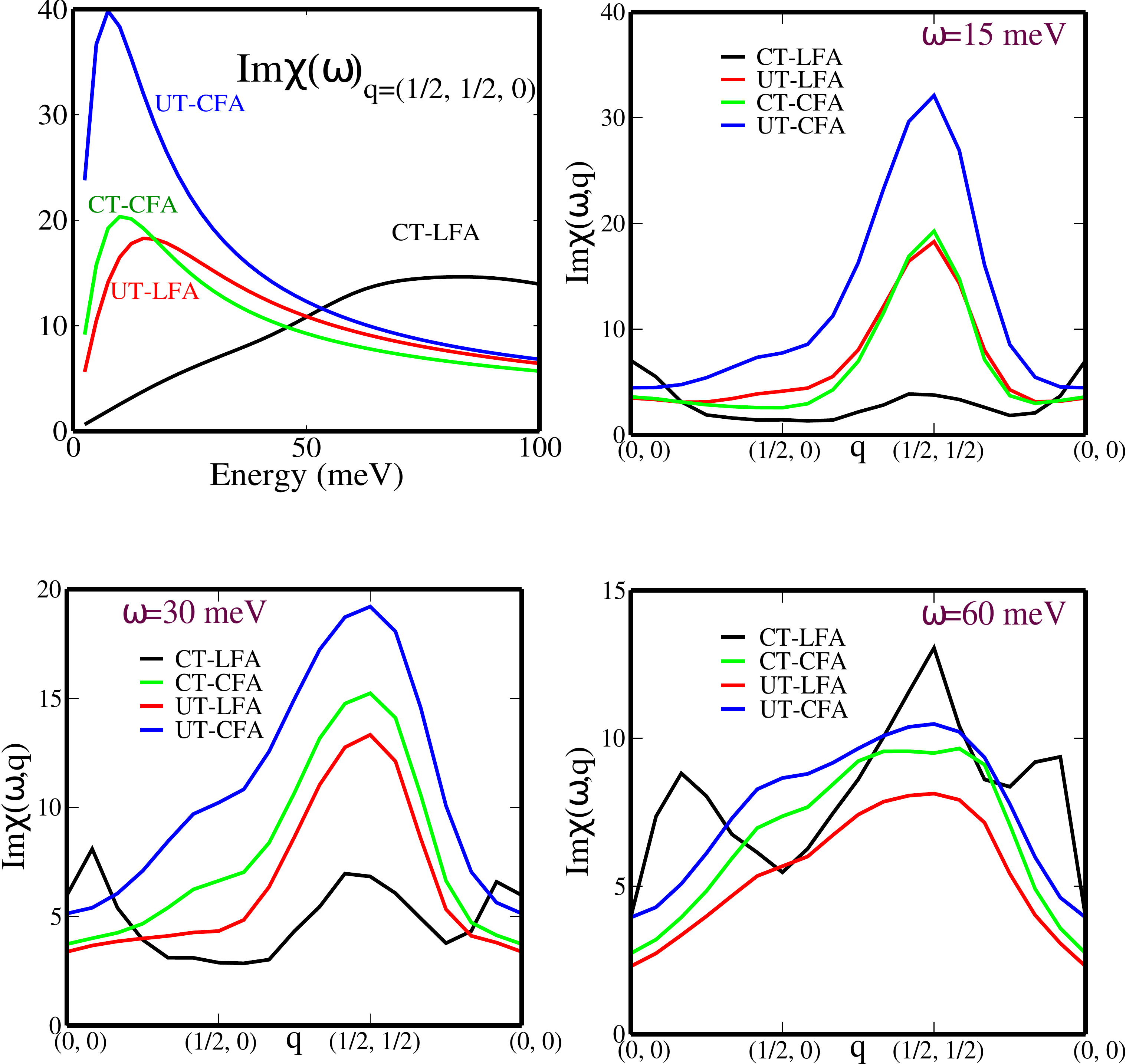}
	\caption{The low energy behavior of Im$\chi(q,\omega)$ is shown for four candidates at
		q=($\frac{1}{2},\frac{1}{2}$,0). The more intense the peak is, higher is the T$_{c}$. Three different energy cuts at
		15, 30 and 60 meV for Im$\chi(q,\omega)$ are resolved along the q-path
		(H,K,L=0)=(0,0)-($\frac{1}{2},0$)-($\frac{1}{2},\frac{1}{2}$)-(0,0) to stress the low-energy concentration of glue in
		the superconducting phases. (The intensity of CT-LFA peak is artificially multiplied by five to bring excitations for
		CT and UT phases to the same scale)}
	\label{fig:intensity}
\end{figure}

\begin{figure}[ht!]
\includegraphics[width=0.49\textwidth]{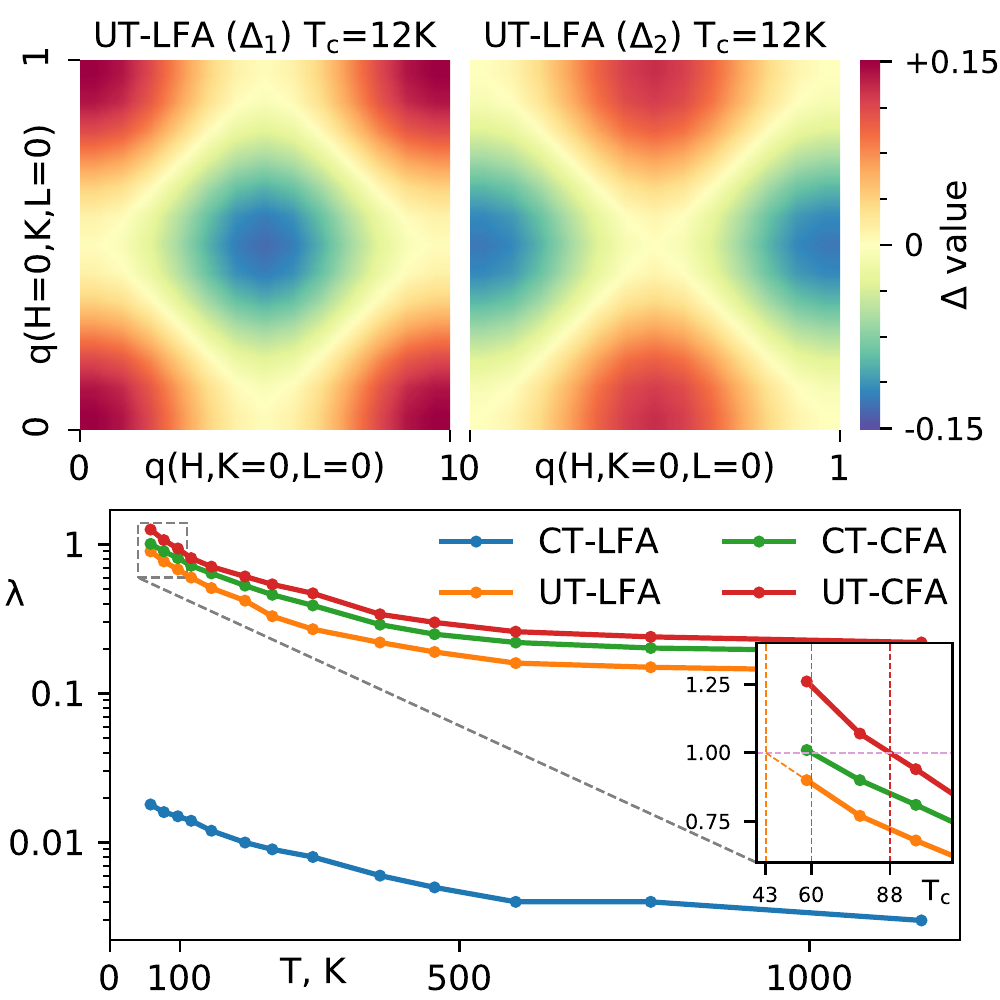}
\caption{The superconducting instability is absent in the CT-LFA phase. The superconducting instability corresponding to
  the leading ($\lambda_{1}$) and lagging ($\lambda_{2}$) eigenvalues of the solutions to the linearized Eliashberg
  equations, $\Delta(q,\omega{=}0)$ are shown for the UT-LFA phase. The evolution of the leading eigenvalue as a
  function of temperature is shown for CT-CFA, UT-CFA and UT-LFA in the bottom panel. In inset we zoom into the low temperature part of the curves to show the estimated T$_{c}$'s.}
\label{gap}
\end{figure}

In Fig.~\ref{fig:intensity} we compare Im$\chi(q,\omega)$ at ($\frac{1}{2},\frac{1}{2}$,0) for four candidates. The UT-CFA has most intense low energy peak followed by CT-CFA and UT-LFA. Low energy spin excitations for CT-LFA is gapped at ($\frac{1}{2},\frac{1}{2}$,0). Further, we take three energy cuts of Im$\chi(q,\omega)$ at $\omega$=15, 30, 60 meV along the path (H,K,L=0)=
  (0,0)-($\frac{1}{2},0$)-($\frac{1}{2},\frac{1}{2}$)-(0,0). At 15 meV, the UT-CFA peak is significantly stronger than
the rest; CT-LFA has weak uniform spin excitation at $q$=0, and is almost entirely suppressed at
($\frac{1}{2},\frac{1}{2}$,0). It appears that an intense low energy peak which is simultaneously broadened in momentum space provides maximum favorable glue for superconducting ordering. For higher energy $\omega$=30 and 60 meV, cuts the sharp difference between UT-CFA and
others start to diminish and the spin excitations for all systems become broad and incoherent and nearly
comparable. CT-LFA shows a clear two-peak structure associated with the high energy paramagnon branch and it disperses to $\sim$500 meV (see SM).     
The eigenvalues and eigenfunctions of superconducting susceptibilities, superconducting pairing symmetries can not be
extracted from the spin dynamics alone. We compute the full two particle scattering amplitude in the particle-particle
channel within our DMFT framework, and we solve Eliashberg equations in the BCS low energy
approximation~\cite{swag19,yin,hyowon_thesis}. We resolve our eigenfunctions of the gap equation into different inter- and
intra-orbital channels, and observe the trend in the leading eigenvalues with temperature in both CT and UT phases. We
observe that there are two dominant eigenvalues of the gap equation. The eigenvalues increase with decreasing T in the
UT-LFA, UT-CFA and CT-CFA, while they are vanishingly small (at least one order of magnitude smaller than the UT phase) and
(in the CT-LFA phase) insensitive to T. The corresponding eigenfunctions in the UT-LFA phase have extended \emph{s}-wave (leading eigenfunction $\Delta_{1}$ for 
eigenvalue $\lambda_{1}$) and d$_{x^2-y^2}$ (lagging eigenfunction $\Delta_{2}$ for eigenvalue $\lambda_{2}$) characters (see Fig.~\ref{gap}). We also find that these instabilities are primarily in
the intra-orbital d$_{xy}$-d$_{xy}$ channel and the inter-orbital components are negligible. In both the UT and CT-CFA
phases the only instability appears to be of extended \emph{s}-wave nature. We track the temperature at which the
superconducting susceptibility diverges (the leading eigenvalue approaches one) to estimate T$_{c}$ (see Fig.~\ref{gap}). We find that the pairing vertex $\Gamma$ rises steeply with lowering temperatures and the leading eigenvalue $\lambda$ follows the temperature dependence of $\Gamma$ (see SM)). Suppression of the charge component of $\Gamma$ leads to no qualitative change to the temperature dependence of $\lambda$ and only weakly changes its magnitude (see SM). Our results suggest that T$_{c}$ is directly proportional to the strength of the low energy peak at ($\frac{1}{2},\frac{1}{2}$), which is further controlled by the correlations and scattering in the Fe-3d$_{xy}$ state.

To conclude, we establish the interplay between the band structure and correlations that lead to emergence (suppression)
of superconductivity in the UT-LFA (CT-LFA) phase.  We establish a direct correspondence between the proximity of the
d$_{xy}$ state to the Fermi energy, and show that it contributes to enhanced low energy scattering and significantly
incoherent quasi-particles.  Incoherence affects two-particle features: the spin susceptibilities also show broad and
intense low energy spin fluctuations centered at ($\frac{1}{2},\frac{1}{2}$). As the phase is quenched, in CT-LFA,
d$_{xy}$ is pushed below $E_{F}$, which causes coherent spectral features to emerge with a broad continuum
of spin excitations.  These do not provide glue conducive for Cooper pair formation.  Our conclusions find further
validation in our calculations in UT and CT phases in CFA. UT-CFA was found to have the most intense low energy
susceptibility peak among the four candidates and is predicted to have the highest, were the superconducting
instability not suppressed by entrant first order structural transition.

This work was supported by the Simons Many-Electron Collaboration.  We acknowledge PRACE for awarding us access to SuperMUC at GCS@LRZ, Germany, STFC Scientific Computing Department's SCARF cluster, Cambridge Tier-2 system operated by the University of Cambridge Research Computing Service (www.hpc.cam.ac.uk) funded by EPSRC Tier-2 capital grant EP/P020259/1.

\pagebreak
\vspace{0.5in}
\pagebreak

\section*{Supplemental Material}
\section*{Crystal structure, quasiparticle properties and susceptibilities} In this supplemental material, we show we list the input structural parameters for our calculations, and the orbitally resolved quasi-particle weight and scattering rates in different compounds, as extracted from QS\emph{GW}+DMFT. We also show the Im$\chi(q,\omega)$ upto 500 meV to demonstrate the itinerant character of spin excitations in CT-LFA.

\begin{table}
	\begin{center}
		\footnotesize
		\begin{tabular}{ccccccccccc}
			\hline
			variants & a (\AA) & c(\AA) & h$_{se}$\\
			\hline
			
			CT-LFA~\cite{akira}   & 4.0035 & 11.0144 & 0.3589 \\
			
			UT-LFA~\cite{akira} & 3.9376 & 11.7317 & 0.3657  \\
			
			CT-CFA~\cite{chen2016} & 3.873 & 11.5647 & 0.3657  \\
			
			UT-CFA~\cite{ni2008} & 3.8915 & 11.69 & 0.372  \\
		\end{tabular}
	\end{center}
	\vbox{\vskip -12pt}
	\caption{Structural parameters, pnictogen height h$_{as}$ as a fraction of $c$ entering as inputs for CT-LFA, UT-LFA, CT-CFA, UT-CFA.}
	\label{tab:strux}
\end{table}

\begin{table}
	\begin{center}
		\footnotesize
		\begin{tabular}{ccccccccccc}
			\hline
			variants & U (eV) & J (eV) \\
			\hline
			
			CT-LFA~\cite{akira}   & 3.9 & 0.72  \\
			
			UT-LFA~\cite{akira} & 3.88 & 0.72  \\
			
			CT-CFA~\cite{chen2016} & 3.97 & 0.73  \\
			
			UT-CFA~\cite{ni2008} & 3.99 & 0.73   \\
		\end{tabular}
	\end{center}
	\vbox{\vskip -12pt}
	\caption{Hubbard parameters U, J computed using QS\emph{GW} + C-RPA  for CT-LFA, UT-LFA, CT-CFA, UT-CFA.}
	\label{tab:strux}
\end{table}

\begin{table}
	\begin{center}
		\footnotesize
		\begin{tabular}{ccccccccccc}
			\hline
			variants & $\Gamma_{x^2-y^2}$ & z$_{x^2-y^2}$ & $\Gamma_{xz,yz}$ & z$_{xz,yz}$ & $\Gamma_{z^2}$ & z$_{z^2}$ & $\Gamma_{xy}$ & z$_{xy}$\\
			\hline
			CT-LFA & 14 & 0.67 & 18 & 0.63 & 12 & 0.67 & 21 & 0.72 \\
			
			UT-LFA & 28 & 0.54 & 60 & 0.54 & 32 & 0.5 & 67 & 0.49\\
			
			UT-CFA & 20 & 0.50 & 45 & 0.43 & 23 & 0.47 & 49 & 0.44 \\
			
			CT-CFA & 26 & 0.53 & 50 & 0.45 & 24 & 0.50 & 47 & 0.47 \\
			
			
			
		\end{tabular}
	\end{center}
	\caption{Quasi-particle renormalization factor (\emph{Z}), single-particle scattering rate ($\gamma$) for CT and UT phases.}
	\label{tab:zandgamma}
\end{table}

\begin{figure}[ht!]
	\includegraphics[width=0.50\textwidth]{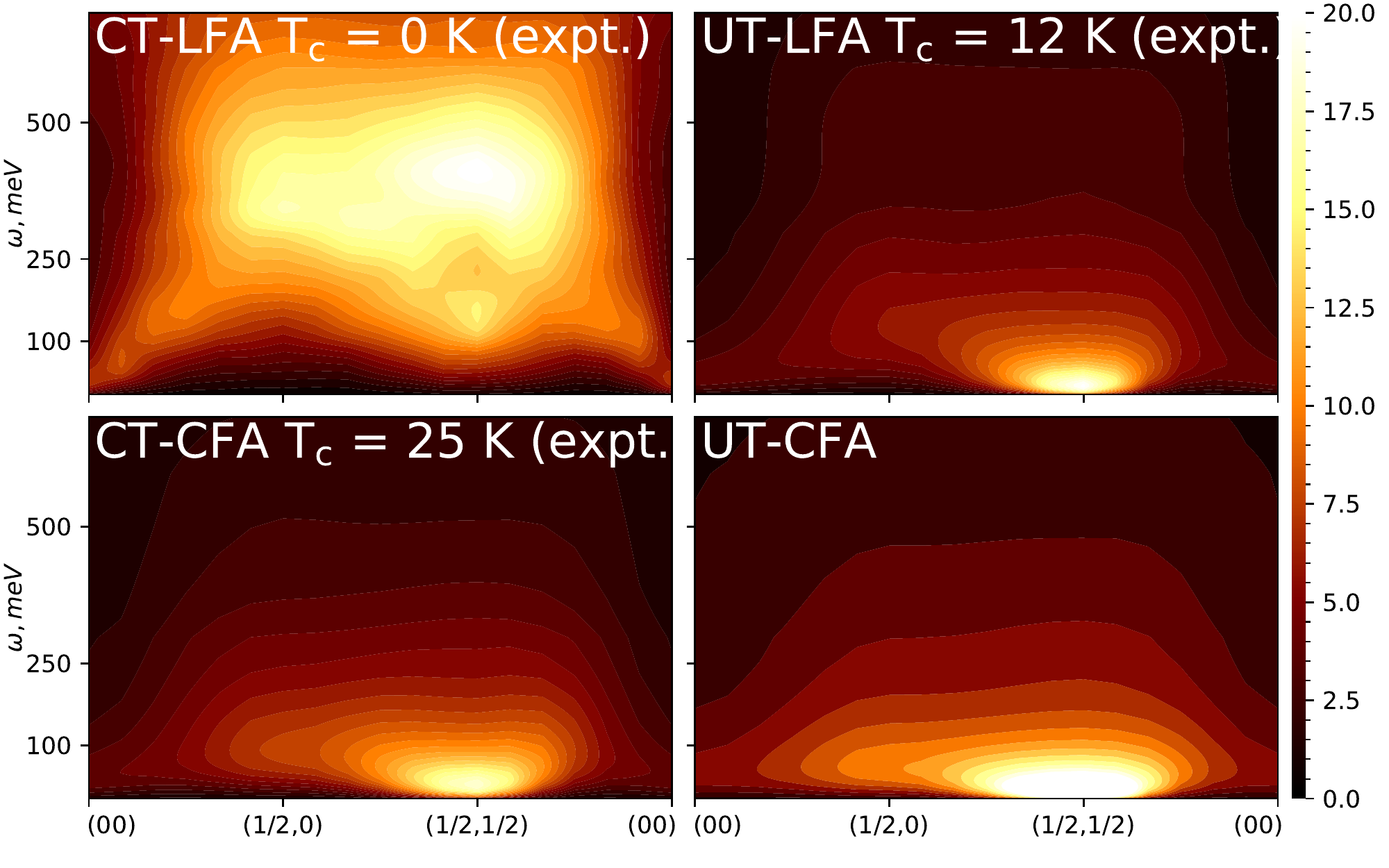}
	\caption{The energy and momentum resolved spin susceptibility Im$\chi(q,\omega)$ (in the top panel from left to right)
		shown for the CT and UT phases respectively over much higher energies to stress the absence of low energy glue in CT-LFA phase, and its nearly band like spin excitation character. The intensity in the CT phase is artificially multiplied by five to bring excitations for CT and UT phases to
		the same scale.}
	\label{fig:chi}
\end{figure}

\vspace{-0.125in}
\subsection*{Note on U and J}

We performed calculations with U and J taken from constrained RPA values for the FeSe and LiFeAs.  Subsequent
constrained RPA calculations on the pnictide considered here (Table II above) indicate that they are about 15\% larger 
than the values we used.  with the change in U and J uniform across all four compounds.  Also the ratio J/U$\sim$0.17
is essentially unchanged.  While correlations should increase, the conclusions will not change since the adjustment is
small and uniform, and and we cannot justify repeating the calculations, in light of the high cost of these
calculations.  

\subsection*{A note on T$_{c}$ estimation} We compute the pairing eigenvalues by solving the linearized Eliashberg equation at different temperatures in the normal phase. The temperature at which the leading eigenvalue becomes one, is where the particle-particle ladder sum (superconducting pairing susceptibility) diverges and corresponds to the T$_{c}$ for that material (the entire method is detailed in our recent work~\cite{swag19} and Park's thesis~\cite{hyowon_thesis}). The local three frequency, orbital dependent vertex functions for solving these linearized Eliashberg equations are computed from CT-QMC. CT-QMC is quantum monte carlo based finite temperature solver. It is fairly expensive to sample the desired vertex functions from CT-QMC, for example, at 300 K, we sample the CT-QMC vertex by launching the calculation on 40,000 cores for 4 hours. 

The superconducting pairing susceptibility $\chi^{p-p}$ is computed by dressing the non-local pairing
polarization bubble $\chi^{0,p-p}(\textbf{k},i\nu)$ with the pairing vertex $\Gamma^{irr,p-p}$ using the
Bethe-Salpeter equation in the particle-particle channel.
\begin{eqnarray}
\chi^{p-p} = \chi^{0,p-p}\cdot[\mathbf{1}+\Gamma^{irr,p-p}\cdot\chi^{0,p-p}]^{-1}
\end{eqnarray}
$\Gamma^{irr,p-p}$ in the singlet (s) channel is obtained from the magnetic (spin) and density (charge) particle-hole reducible vertices by
\begin{eqnarray} 
&\Gamma_{{\alpha_{2},\alpha_{4}\atop \alpha_{1},\alpha_{3}}}^{irr,p-p,s}(\textbf{k},i\nu,\textbf{k}',i\nu') = \Gamma_{{\alpha_{2},\alpha_{4}\atop \alpha_{1},\alpha_{3}}}^{f-irr}(i\nu,i\nu')\nonumber\\
&+\frac{1}{2}[\frac{3}{2}\widetilde{\Gamma}^{p-h,(m)}\nonumber\\
&-\frac{1}{2}\widetilde{\Gamma}^{p-h,(d)}]_{{\alpha_{2},\alpha_{3}\atop \alpha_{1},\alpha_{4}}}(i\nu,-i\nu')_{\textbf{k}'-\textbf{k},i\nu'-i\nu}\nonumber\\
&+\frac{1}{2}[\frac{3}{2}\widetilde{\Gamma}^{p-h,(m)}\nonumber\\
&-\frac{1}{2}\widetilde{\Gamma}^{p-h,(d)}]_{{\alpha_{4},\alpha_{3}\atop \alpha_{1},\alpha_{2}}}(i\nu,i\nu')_{-\textbf{k}'-\textbf{k},-i\nu'-i\nu}
\label{eq:Gamma_pp_nonloc1}
\end{eqnarray}

Finally, $\chi^{p-p}$ can be represented in terms of eigenvalues $\lambda$ and eigenfunctions $\phi^{\lambda}$
of the Hermitian particle-particle pairing matrix.
\begin{eqnarray}
\chi^{p-p}(k,k') & = & \sum_{\lambda}\frac{1}{1-\lambda}\cdot(\sqrt{\chi^{0,p-p}(k)}\cdot\phi^{\lambda}(k))\nonumber\\
&\cdot(&\sqrt{\chi^{0,p-p}(k')}\cdot\phi^{\lambda}(k'))
\end{eqnarray}
The pairing susceptibility diverges when the leading eigenvalue approaches unity. 
When the particle-particle ladder sum $\chi^{p-p}$ = $((\chi^{0,p-p})^{-1}-\Gamma^{p-p})^{-1}$ diverges, the normal state becomes unstable towards
superconductivity. In a temperature dependent calculation this corresponds to the T$_{c}$ where the leading eigenvalue of the matrix $\Gamma^{p-p}\chi^{0}$ reaches one (as shown in Eqn. (3)). The
eigenvector corresponding to the leading eigenvalue $\lambda$ gives the symmetry of the superconducting order parameter $\Delta_{\alpha,\beta}(k,\nu)$. In an ideal scenario we need 
need to solve the eigenvalue problem of the following matrix;

\begin{eqnarray}
&-K_{B}T\sum_{k'\nu'\alpha'\beta'\gamma\delta} \gamma^{p-p,s}(\alpha\beta k \nu; \alpha'\beta'k'\nu')\chi^{0,p-p}_{\alpha'\beta'\gamma\delta}(k',\nu')\nonumber\\
&\Delta(\gamma\delta)(k',\nu')=\lambda\Delta_{\alpha,\beta}(k,\nu)
\end{eqnarray}

\begin{figure}[ht!]
	\includegraphics[width=0.50\textwidth]{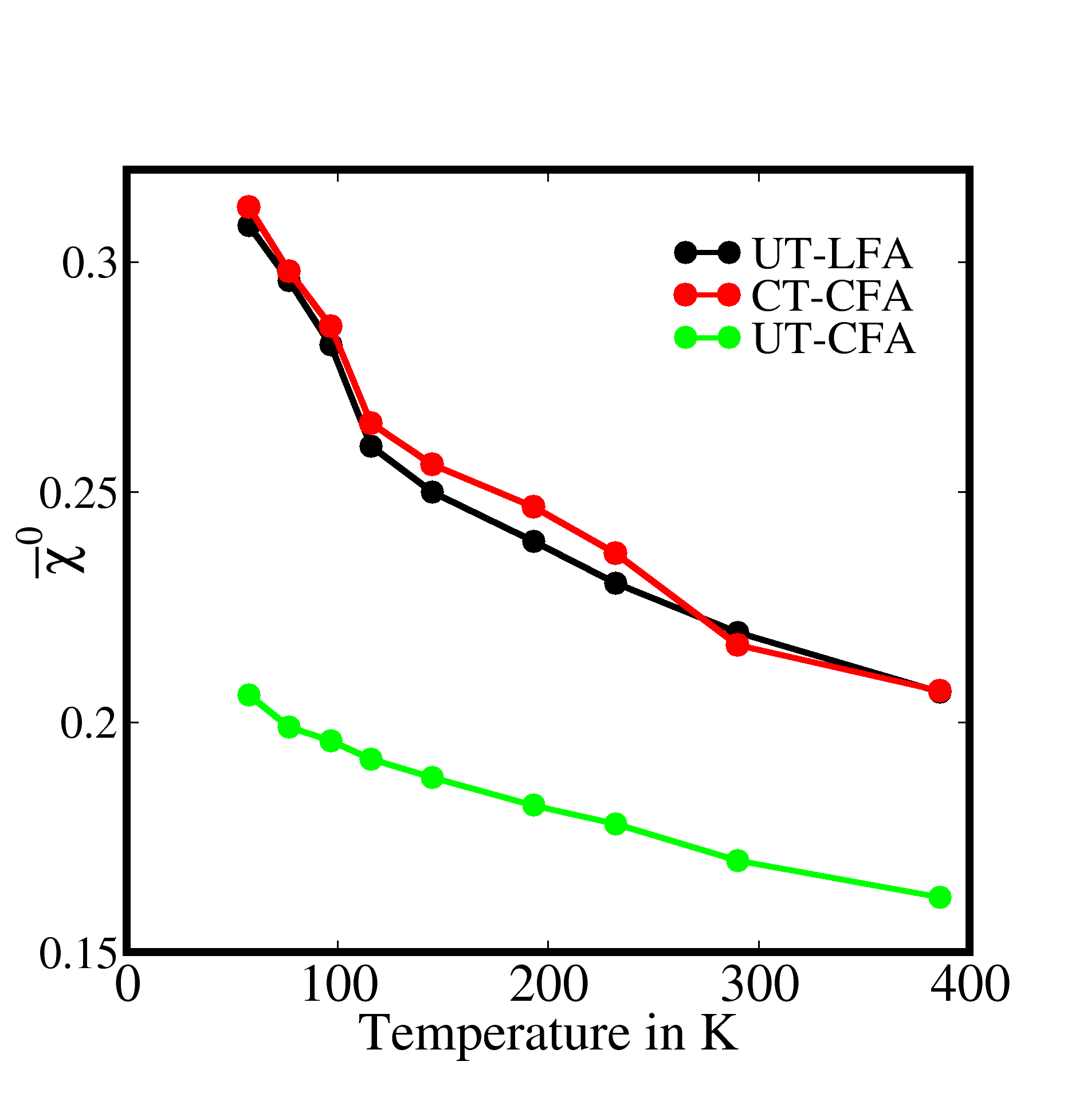}
	\includegraphics[width=0.50\textwidth]{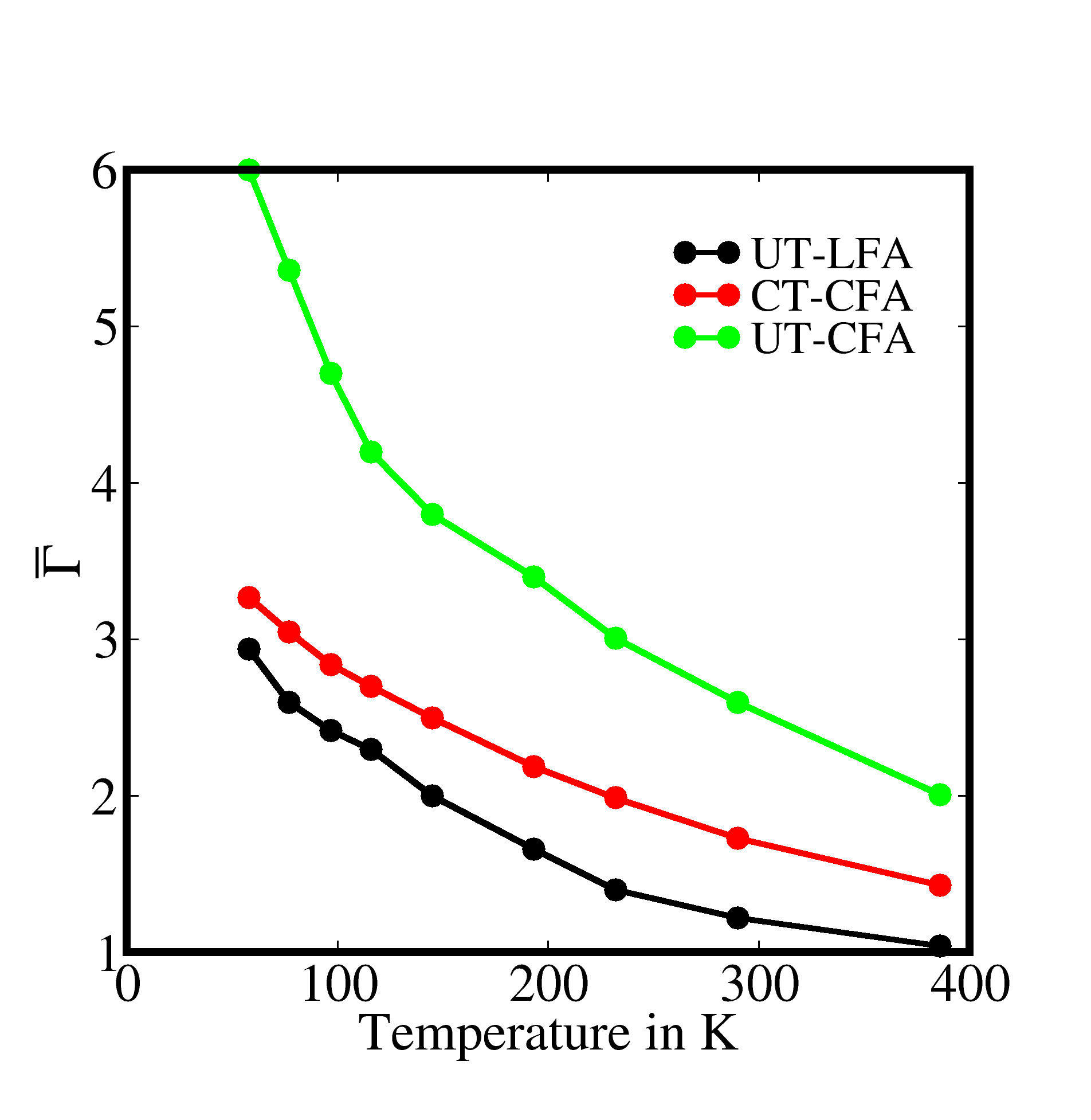} 	
	\caption{The particle-particle bubble $\bar\chi^0$ and the particle-particle interaction vertex $\bar\Gamma$ (projected onto the leading pairing symmetry channel) are plotted as functions of temperature. The leading eigenvalue of the Eliashberg gap equation $\lambda$ follows primarily the steep rise in $\bar\Gamma$ with lowering temperatures.}
	\label{fig:chi}
\end{figure}

\begin{figure}[ht!]
	\includegraphics[width=0.50\textwidth]{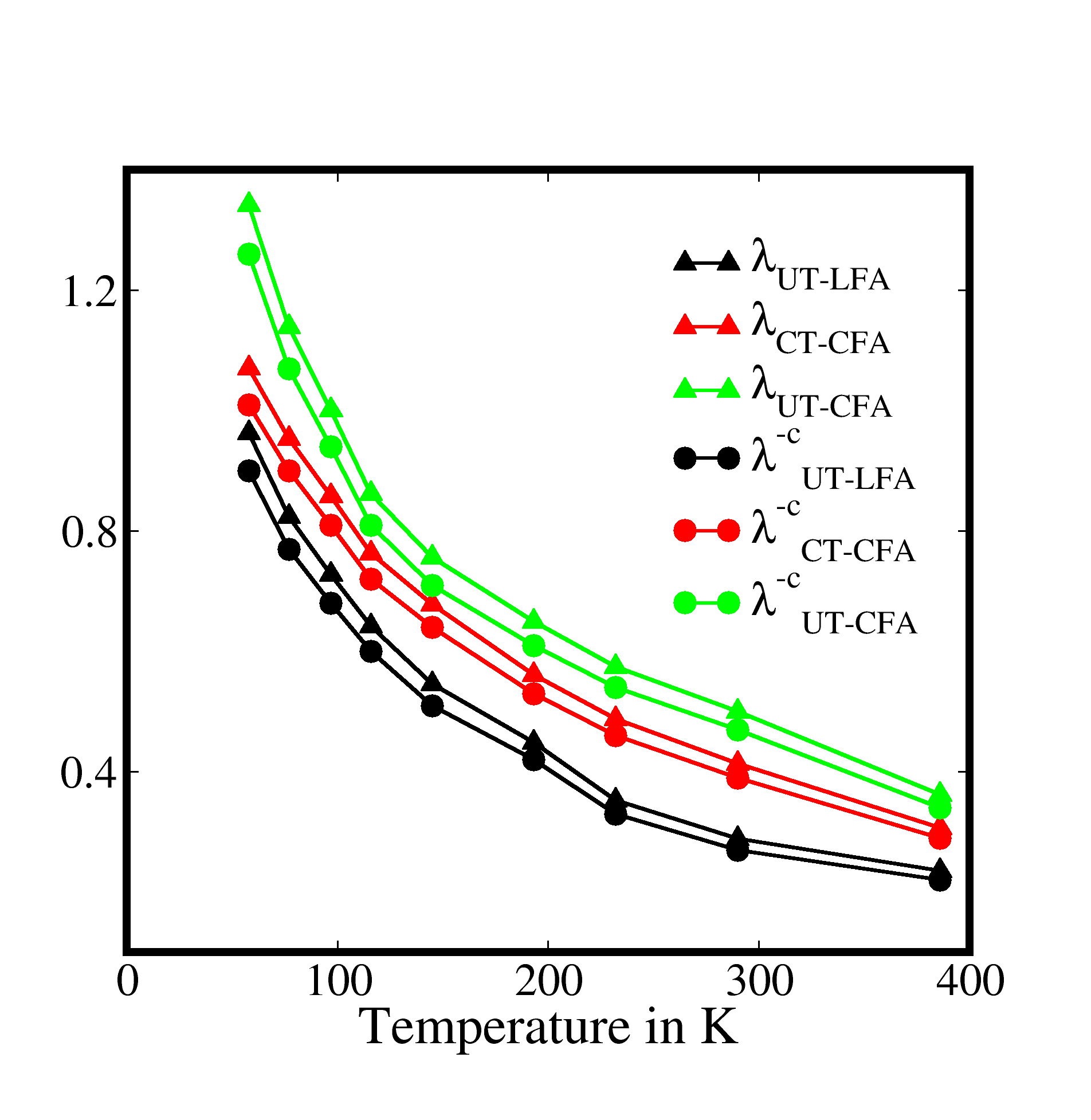}
	\caption{The leading eigenvalue of the Eliashberg gap equation $\lambda$ is shown to get very weakly affected once the charge component of the pairing interaction vertex $\Gamma$ is suppressed (see Eqn.(2)). The original $\lambda$ with both spin and charge components kept in the vertex and the one without the charge component $\lambda^{-c}$ have similar temperature dependence.}
	\label{fig:chi}
\end{figure}

The matrix that needs to be diagonalized has a size $(norb^{2}*nomega*nkp)*(norb^{2}*nomega*nkp)$. In case of our materials, we find that even at $\beta$=20, which is roughly T = 580 K, the matrix size that we need to diagonalize is of the size (norb=5, nomega=200, nkp=1000) size (25*200*1000)*(25*200*1000). So we employ BCS low energy approximation to diagonalize the matrices at different temperatures and extract the eigenvalue spectrum. The BCS approximation amounts to using the $\Gamma^{p-p}$ strictly from the lowest energy ($\nu=0^+$, $\nu'=0^+$, $\omega$=0). However, $\Gamma^{p-p}$ contains all relevant momentum and orbital structure, and bubble contains information from all energies, momentum and orbitals. This appears to be reasonable approximation as the vertex contains essential features of superconductivity which is a low energy phenomenon. Additionally the pairing vertex also shows the essential temperature dependent enhancement that is prerequisite to Cooper pairing. We project the computed bubble and vertex functions on to the leading superconducting pairing symmetry channel $\Delta^{k}$$\sim (cosk_{x}+cosk_{y}$) and show their temperature dependent behaviour.

\begin{eqnarray}
\bar\Gamma=\frac{\sum_{k,k'}\Delta^k\Gamma(k,k')\Delta^{k'}}{\sum_{k}(\Delta(k))^2}
\end{eqnarray}

\begin{eqnarray}
\bar\chi^0=\frac{\sum_{k,k'}\Delta^k\chi^0(k,k')\Delta^{k'}}{\sum_{k}(\Delta(k))^2}
\end{eqnarray}

Further, we suppress the charge (density) component of the $\Gamma^{p-p}$ to show that the eigenvalues ($\lambda^{-c}$) only get very weakly affected. This is in complete consistency with what we show in the main paper that the superconducting instability is in one-to-one correspondence with the spin instability and the pairing is primarily mediated via spin fluctuations. 


%

\end{document}